\documentclass[prb,twocolumn,showpacs]{revtex4}

\usepackage{graphicx}
\usepackage{dcolumn}
\usepackage{bm}
\usepackage{color}

\def\Sci{Science}
\def\naturemat{Nature Mater.}
\def\SSC{Solid State Commun.}

\def\PRL{Phys. Rev. Lett. }

\def\PRB{Phys. Rev. B }

\def\N{Nature}
\def\NP{Nature Phys.}

\def\JPSJ{J. Phys. Soc. Jpn.}
\def\APL{Appl. Phys. Lett.}
\def\NN{Nature Nanotech.}

\begin{document}

\title{Raman spectra of bilayer graphene to probe the electrostatic environment}

\author{\hspace{-1.7cm} Paola Gava, Michele Lazzeri,
A. Marco Saitta, and Francesco Mauri\\[1em]  
IMPMC, Universit\'es Paris 6 et 7, CNRS, IPGP, 140 rue de Lourmel,
75015 Paris, France
\normalsize
}

\date{\today}

\begin{abstract}
The Raman shift, broadening, and relative Raman intensities of bilayer 
graphene are computed as functions of the electron concentration. 
We include dynamic effects for the
phonon frequencies and we consider the gap induced in the band 
structure of bilayer graphene by an external electric field. 
We show that from the analysis of the Raman spectra of gated bilayer
 graphene it is possible to quantitatively identify the amount of 
charges coming from the atmosphere and from the substrate. 
These findings suggest that
Raman spectroscopy of bilayer graphene can be used to 
characterize the electrostatic environment of few-layers graphene.
\end{abstract}


\maketitle

\section{introduction}

Graphene-based systems have recently attracted much attention 
from both the experimental and the theoretical point of view.
Graphene is in fact characterized by a high carrier-mobility, 
\cite{Novoselov,Novoselov-1,Zhang} which make those systems 
very exciting in view of future applications in nanoelectronics.
In standard experimental setups, few-layers graphene (FLG) interacts with the environment 
through doping charges coming from the top ($n_{top}$) and
from the bottom ($n_{bot}$) of the system.
These charges determine the external electric field  
$\left[E = (n_{top}-n_{bot})|e|/(2\epsilon_0)\right]$
and the total electron concentration ($n=n_{top}+n_{bot}$).
$E$ and $n$ can thus be varied independently changing the charges from the two sides.
$n_{top}$ and $n_{bot}$ can be intentionally induced
by applying gate voltage differences between the system and the substrate,
 or via atoms/molecules deposition on top of the system.
On the other hand, important unintentional doping charges are typically 
present.
For instance, in FLG obtained by exfoliation on ${\rm SiO_2}$
\cite{Novoselov,Novoselov-1,Zhang,Oostinga-Nature,Zhang-Nature,Castroneto_PRL,Ferrari,Pisana,Casiraghi,Stampfer,Das,Yan-PRL-98,Das-1,Yan-PRL-101,Pimenta-phonon,Kuzmenko-infrared,Tang-arxiv,Kuzmenko-arxiv,Zhang-LM} or epitaxially 
grown on SiC, \cite{Ohta_science} a charge transfer occurs between 
the substrate and the system, giving rise to a finite $n_{bot}$. 
In analogy, an additional $n_{top}$ can be accidentally induced by
the uncontrolled adsorption of molecules from the atmosphere.

Among FLG, the bilayer graphene is particularly interesting because it becomes a tunable
band-gap semiconductor under the application of an electric field
perpendicular to the system.
\cite{Ohta_science,Oostinga-Nature,Zhang-Nature,Castroneto_PRL,McCann-PRB-74,Min-PRB-75,Aoki-solstatecomm-142,Gava}
In this context, the determination of the electric field is crucial in order to control the band-gap.
Moreover, in graphene charge impurities originating from the top or from the substrate 
are the main source of scattering which reduces conduction performances. \cite{Chen}
Therefore, the determination of the charge transfer from the substrate or from the
atmosphere is
highly desirable for possible applications.
Although experimental techniques allow to estimate
the total electron concentration on the system, an accurate determination
of the respective values of top and bottom charges has never been achieved so far,
and can be particularly challenging, for instance,
when doping is intentionally induced by deposition of molecules or polymeric electrolyte.
In this work we use a tight binding (TB) model fitted on 
ab initio calculations to compute the Raman shift, broadening, and relative
Raman intensity of the $G$ modes in bilayer graphene, as a function of the electron concentration,
for different values of top charges. 
In particular, the screening properties of the system in presence of
an external electric field are described using ab initio density functional theory calculations (DFT),
including the $GW$ correction,
while a TB model is used to reproduce
the DFT calculated, $GW$ corrected, band structure in the full Brillouin zone.
We show that from the measured Raman spectra of bilayer graphene
it is possible to determine the external charge distribution  
and thus the external electric field and the actual carriers concentration.
This result is especially relevant since it shows that Raman spectroscopy,
which is already widely used to investigate graphene-based systems,
can be used to characterize the electrostatic environment of the sample. 
Moreover, since the charges coming from the atmosphere and the substrate
are not expected to depend on the number of layers,
Raman measurements on bilayer graphene can also be used to determine 
the origin and the amount of the unintentional doping
of monolayer and few-layers graphene in the same environment.

\begin{figure}[t]
  \centering
  \includegraphics[width=0.9\columnwidth]{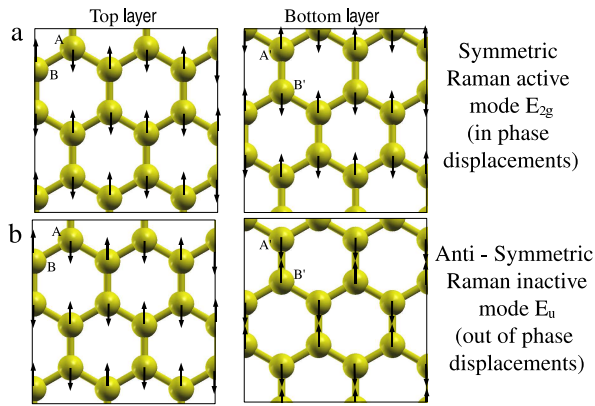}
  \caption{ (Color online) Schematic representation of the Raman active mode $E_{2g}$ (a) and the Raman inactive mode $E_u$ (b) in bilayer graphene. A,B are the inequivalent carbon atoms in the top layer, A' and B' are the inequivalent carbon atoms in the bottom layer.
In a Bernal stacking configuration, A and A' are the superposed atoms.
 }
  \label{Raman-IR}
\end{figure}

The Raman spectra of monolayer graphene are characterized by a 
doubly-degenerate $G$ peak ($E_{2g}$ mode) at around 1580 ${\rm cm^{-1}}$.
\cite{Ferrari,Pisana,Casiraghi,Stampfer,Das,Yan-PRL-98,Das-1}
This mode shows a strong electron-phonon coupling, 
which induces a phonon renormalization when $n$ is varied.
Therefore, Raman can be used to measure the total electron concentration.
In bilayer graphene, in the absence of an external electric field
the $G$ peak splits, as in graphite, in a doubly-degenerate 
Raman active mode ($E_{2g}$) and a doubly-degenerate, Raman inactive, infra-red active mode ($E_u$).
The $E_{2g}$ mode is characterized by a symmetric in-phase displacement 
of the atoms in the two layers (Fig.\ref{Raman-IR}-a), 
whereas $E_u$ is characterized by an antisymmetric out-of-phase 
displacement of those atoms (Fig.\ref{Raman-IR}-b).
Most Raman measurements on bilayer graphene show a single $G$ peak 
whose position is used, as in monolayer graphene, to measure the 
electron concentration on the system. \cite{Yan-PRL-101,Das}
Interestingly, the splitting of the $G$ peak has 
been observed in the Raman spectra of bilayer graphene, \cite{Pimenta-phonon}
and it has been recently attributed to symmetry breaking
due to the application of an external electric field.\cite{Ando-new}
Moreover, other experimental works recently 
reported on the infra-red spectra of gated bilayer graphene.
\cite{Zhang-Nature,Kuzmenko-infrared,Tang-arxiv,Kuzmenko-arxiv,Zhang-LM}
These findings suggest that a deep understanding of the 
behavior of a splitted $G$ mode  
in gated bilayer graphene could lead to quantify not only 
the total electron concentration $n$ but also the separate values 
of $n_{top}$ and $n_{bot}$. 

\section{Computational details}

In order to calculate the electron-phonon coupling (EPC) component of the phonon frequencies, 
broadenings, and relative Raman intensities as functions of $n$ and $n_{top}$, 
we consider the $\Gamma$ phonon self-energy, \cite{PhysRevB.6.2577,PhysRevB.9.4733}
projected onto the subspace of the two $E_{2g}$ and the two $E_u$ modes:
\begin{equation}
\Pi_{\mu \nu}(n_{top},n) = \frac{\hbar}{M\omega_0 N_k} \sum_{{\bf k},i,j} \frac{D^{\mu}_{ji} D^{\nu}_{ij}\ (f_{{\bf k}i}-f_{{\bf k}j})} {\epsilon_{{\bf k}i}-\epsilon_{{\bf k}j}+\hbar \omega_0 + {\rm i} \eta},
\label{EPC}
\end{equation} 
where the sum is on the electron wave vector ${\bf k}$ and 
the electronic $\pi$ bands $i,j$. $\mu,\nu$=1,4 are the 
phonon indexes, $N_k$ is the number of ${\bf k}$ vectors, 
and $f_{{\bf k}i}$ is the occupation of the electron state 
$\arrowvert {\bf k}i\rangle$ with energy $\epsilon_{{\bf k}i}$.
$D^{\mu}_{ij} = \langle{\bf k}i\arrowvert \Delta H^{\mu}\arrowvert{\bf k}j\rangle$ 
is the EPC and $ \Delta H^{\mu}$ is the Hamiltonian derivative with respect 
to the atomic displacement corresponding to the $\mu$ phonon.
 $\eta$ equals 0.009 eV and M is the atomic mass.

From the phonon self-energy one can calculate only frequency 
variations due to changes in the electron concentration and in the band structure.
In order to obtain the absolute frequencies  
we use the following 4${\rm \times}4$ matrix:
\begin{eqnarray}
\Omega_{\mu \nu}(n_{top},n)&=&\left(\omega_0 + \Delta \omega(n)- \Pi_0 + {\rm i}\frac{\gamma_{an}}{2} \right) \delta_{\mu \nu} \nonumber  \\
&+& \Pi_{\mu \nu}(n_{top},n), 
\label{omega}
\end{eqnarray} 
where $\omega_0$=1581.5 ${\rm cm^{-1}}$ is the experimentally measured  
frequency of the Raman active $G$ peak in bilayer graphene, in absence of doping and electric field.
\cite{Yan-PRL-101} 
$\gamma_{an}$ is the contribution 
to the broadening of the $G$ modes in graphene and graphite
from the anharmonic phonon-phonon interaction, whose value is estimated to be 1.8 ${\rm cm^{-1}}$.
\cite{Bonini}
$\Pi_0$ is the phonon self-energy of the doubly degenerate $E_{2g}$ mode 
calculated for $n=0$ and $n_{top}=0$, and it is given by
$\Pi_0 = {\rm Re}\left[ u^{k}_{\mu} \Pi_{\mu \nu}(0,0) u^{k}_{\nu} \right]$,
where $k=1,2$ corresponds to the two $E_{2g}$ modes and 
$u^{k}_{\mu}$ are the corresponding eigenvectors.
In presence of doping charge the lattice parameter changes and
the corresponding variation of the $G$ modes frequencies 
can be obtained for zero electric field by 
ab initio calculations:\cite{Lazzeri-Mauri}
$\Delta \omega(n) = -5.75\  10^{3}\ \Delta a(n)$~cm$^{-1}$,
where $\Delta a(n)$ is the relative variation of the lattice parameter,
as in Eq.(2) of Ref.\cite{Lazzeri-Mauri}
In this work we do not include in the phonon
calculations the effects due to direct interaction of the system with adsorbates
and with the substrate donating doping charge. However, as shown for intercalated
graphite,\cite{Boeri,Calandra} we assume that this is not relevant also for
the in-plane vibrational modes of bilayer graphene.

The eigenvalues of $\Omega_{\mu \nu}$ are of the form 
$(\omega_i+{\rm i} \gamma_i /2)$, where $\omega_i$ is the frequency of 
phonon $i$ and $\gamma_i$ is the full width half maximum (FWHM),
given by the EPC and the anharmonic phonon-phonon interaction.
In the general case of finite $n$ and $n_{top}$ 
the four eigenmodes of $\Omega_{\mu \nu}$, 
$\varepsilon^{i}_{\mu}$, are still two by two degenerate, 
but they are a superposition of the $E_{2g}$ and $E_{u}$ eigenmodes of the
unbiased bilayer graphene. Their relative Raman intensities $I_{R}^{i}$ are calculated as:
\begin{equation}
I_{R}^i = \frac{\sum_{k=1,2} | \varepsilon^{i}_{\mu} \cdot  u^{k}_{\mu}|^2}{\sum_{i}\sum_{k=1,2} | \varepsilon^{i}_{\mu} \cdot  u^{k}_{\mu}|^2},
\label{intensity}
\end{equation}
where $\sum_{i} I_{R}^i$=1.

We have shown that the screening properties of bilayer graphene 
under the application of an external electric field
are characterized by inter- and intra-layer polarizations.\cite{Gava}
Most of the calculations of the band gap as a function of $n$ and $n_{top}$
are based on TB models,\cite{McCann,Castroneto_PRL} which usually consider only the
inter-layer palarization, resulting in an
overestimation of the band gap.
In the present work the band structure of the $\pi$ electrons
[$\epsilon_{{\bf k}i}$ and $\arrowvert {\bf k}i\rangle$
in Eq.(\ref{EPC})] is obtained using the scheme presented in Ref.\cite{Gava}
The band gap 
is computed by ab initio DFT calculations, including the $GW$ corrections,
and both inter- and intra-layer polarizations 
are fully taken into account.  
The full band structure of gated bilayer graphene
is then computed using a TB model, which is able to reproduce
all the important features of the DFT calculated, $GW$ corrected bands, 
including the electron-hole asymmetry.

In order to compute $ \Delta H^{\mu}$ one needs to calculate the 
derivative of the tight binding Hamiltonian with respect to the 
atomic positions. However, only the variation of the first 
nearest-neighbors in-plane hopping parameters $\gamma_1$ turns 
out to be relevant, and this only term is thus included in $ \Delta H^{\mu}$.
The value we use for this quantity 
is 5.8 eV ${\rm \AA^{-1}}$, 
which derives from the ab initio GW-calculated EPC at $\Gamma$ 
for the $E_{2g}$ mode in monolayer graphene. \cite{Lazzeri-Att}

\section{Results}

\begin{figure}
  \centering
\includegraphics[width=0.8\columnwidth]{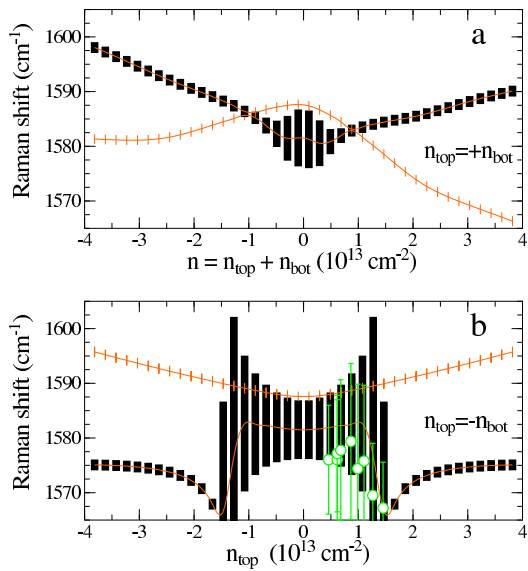}
\caption{(Color online) Raman shift in bilayer graphene for
$n_{top}=n_{bot}$ and for $n_{top}=-n_{bot}$.
Calculated values of the shifts are connected by lines.
For a given value of $n$ (panel a) or $n_{top}$ (panel b) there are two
phonon modes represented with two rectangles.
The height of the rectangles is the FWHM and the areas are proportional
to the relative Raman intensities (i.e. the integrated area of each peak)
of the two modes.
Thus, the ratio of the widths of the two rectangles is equal to the
ratio of the maximum heights of the two Raman peaks.
When the ratio is less than 0.1, the mode with the smallest intensity
is colored in gray (red), otherwise is black.
Circles are experimental results from Ref.\cite{Tang-arxiv}
and the errorbars represent the experimental FWHM.
}
\label{shift}
\end{figure}

\begin{figure}
\centering
\includegraphics[width=0.8\columnwidth]{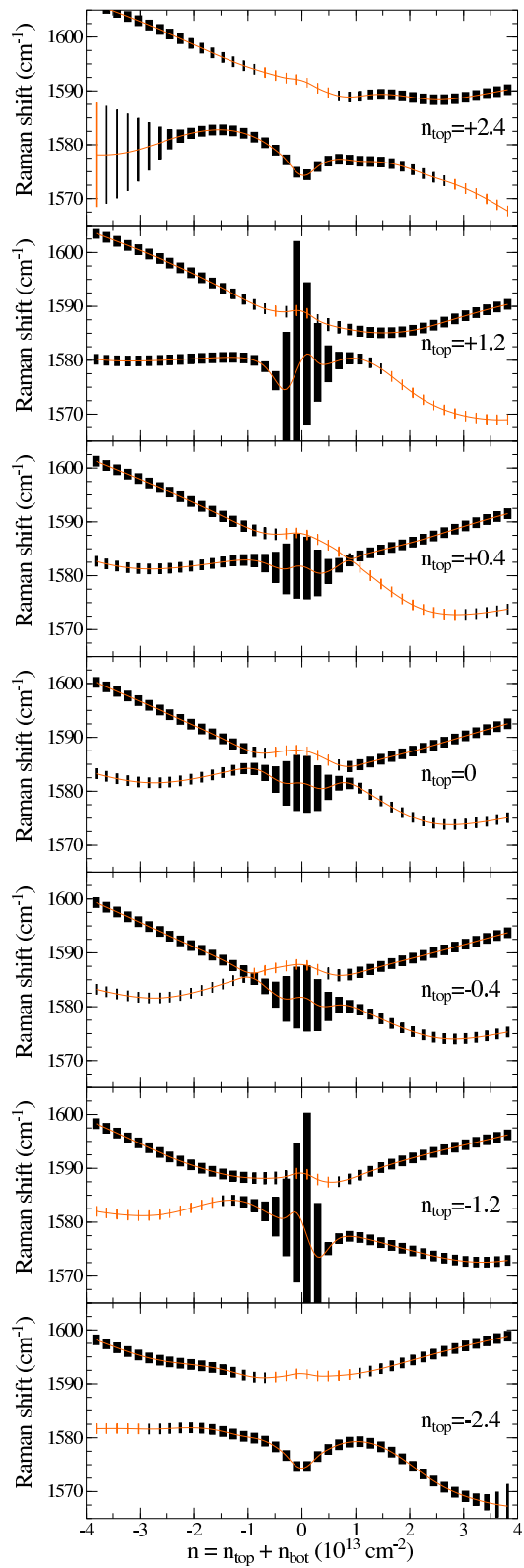} 
\caption{(Color online) Raman shift in bilayer graphene as a function of the
electron concentration $n$, for different values of $n_{top}$.
See the caption of Fig.\ref{shift}.
}
\label{shift-1}
\end{figure}

In Fig.\ref{shift}-a we show the calculated Raman shift 
as a function of $n$,
for the case $n_{top}$=$n_{bot}$,
where the external electric field and the band-gap are kept fixed to zero,
as theoretically studied in Ref.\cite{Ando}
In this case the $E_{2g}$ and $E_{u}$ modes do not mix by symmetry.
Analogously to monolayer graphene, the Raman active modes show 
a singularity when the Fermi energy is half of the phonon energy. 
In Fig.\ref{shift}-b we show the calculated Raman shifts for the case $n_{top}=-n_{bot}$,
as a function of $n_{top}$.
This is a special situation where the external electric field is
varied while $n$ is kept fixed to zero, as realized in recent infra-red experiments.
\cite{Zhang-Nature,Tang-arxiv}
In this case, the mixing between $E_{2g}$ and $E_{u}$ modes is weak
due to the antisymmetric allocation of charges in the two layers.
The Raman active modes show a singularity in the frequency 
and a divergence in the FWHM when the band-gap is of the 
order of the phonon energy.

In the most general situation the electric field and $n$ are both finite. 
In Fig.\ref{shift-1} we show the Raman shift
of bilayer graphene as functions of $n$, for different values of $n_{top}$.
In these cases, the $E_{2g}$ and $E_{u}$ modes do mix, and at certain values of $n_{top}$ and $n$ 
two modes become Raman visible.
Our results show an asymmetry between positive and negative values of $n_{top}$.
For instance, in the case $n_{top}$ = 2.4 ${\rm 10^{13} cm^{-2}}$ the FWHM 
of the lowest mode at $n \approx$ -4 ${\rm 10^{13} cm^{-2}}$ is higher 
than the same quantity for the case $n_{top}$ = -2.4 ${\rm 10^{13} cm^{-2}}$
at $n \approx$ 4 ${\rm 10^{13} cm^{-2}}$. 
This is due to the electron-hole asymmetry in the band structure, which 
is properly described in our calculations.
Moreover, the asymmetry between positive and negative $n$ 
 is enhanced by the effect of the lattice spacing variation induced
by the doping charge. 
Our results are qualitatively in agreement with recent calculations,\cite{Ando-new} 
based on TB model. 
They are however quantitatively different,  
because in our calculations we include the electron-hole asymmetry, 
the lattice spacing variation due to doping charge,
and both inter- and intra-layer polarizations.
The dependence on $n$ of the frequencies,
FWHM, and relative Raman intensities is strongly influenced by  
$n_{top}$, and on the basis of this observation we claim that the amount of uncontrolled 
$n_{top}$ and $n_{bot}$ can be estimated from the Raman spectra 
of bilayer graphene when two modes are observed.

\begin{figure}
\centering
\includegraphics[width=0.9\columnwidth]{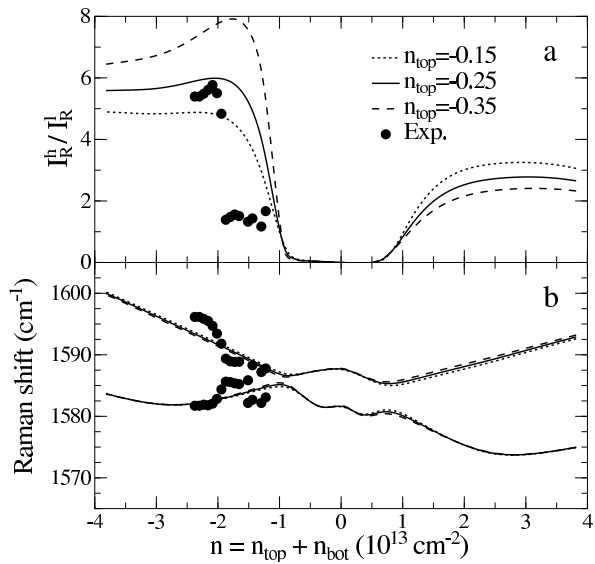}
\caption{Ratio between the relative Raman intensities
of the highest and lowest mode, a, and Raman shift, b, as a function of $n$.
Filled dots are experimental results
from Ref.\cite{Pimenta-phonon,fit},
shifted by $n^0$=-1.8 ${\rm 10^{13} cm^{-2}}$.
Lines are the theoretical values
for $n_{top}$=-0.15, -0.25, -0.35 ${\rm 10^{13} cm^{-2}}$.
}
\label{exp}
\end{figure}

We now compare our calculations to the experimental Raman spectra of 
bilayer graphene where the splitting of the $G$ mode is reported. \cite{Pimenta-phonon}
In this work charges are intentionally induced on the system
by applying a gate voltage between the bilayer and the ${\rm SiO_2}$ substrate. 
However, unintentional $n_{top}$ 
and $n_{bot}$ arising from the atmosphere and the substrate can be present
at zero gate voltage.
By comparing the experimental and calculated Raman shifts
as a function of $n$ for different $n_{top}$,
we estimate a total electron concentration at zero
gate voltage $n^0$ = -1.8 ${\rm 10^{13}cm^{-2}}$. 
In Fig.\ref{exp}-a and -b we show the ratio between the relative Raman
intensities of the highest and lowest mode, and the Raman shifts, respectively, as a function of
the electron concentration $n$, for different values of $n_{top}$.
The former one strongly depends on $n_{top}$,
while the frequency shifts have a weaker dependence.
The best agreement between theory and experiments indicates an unintentional charge 
coming from the atmosphere 
$n^{0}_{top}$=-0.25 ${\rm 10^{13} cm^{-2}}$.
From our estimate of $n^0$,
we deduce an unintentional charge from the substrate
$n^{0}_{bot}$=$n^0$-$n^{0}_{top}$= -1.55 ${\rm 10^{13} cm^{-2}}$.
The agreement between experimental data and theoretical
results is good. However, we notice that the slope of the theoretical curves
is underestimated with respect to the experimental ones. This could be possibly due
to local desorption of molecules and doping variation induced by the
laser light, or to hysteresis effects in the doping dependence on gate voltage.

Finally, in the right side of Fig.\ref{shift}-b, we compare our theoretical 
results to the experimental frequencies 
and broadenings from recent infra-red measurements in Ref.\cite{Tang-arxiv},
where the doping charge is kept fixed to zero and the electric field is varied.
The agreement is excellent with our lower frequency mode.
In our calculations, the lower mode has a weak projection on $E_u$.
However, this mode is strongly coupled
with the electrons, as testified by the large FWHM.
Such coupling enhances the effective charges associated with $E_u$
and increases the infra-red activity. \cite{Kuzmenko-arxiv,Rice-PRL,Rice-PRB}
Indeed, in Fig.3 of Ref.\cite{Tang-arxiv} 
the measured infra-red intensity is maximum
when the band-gap equals the phonon energy (about 0.2 eV), 
i.e. when the FWHM and thus the 
coupling of the mode with the electrons are maximum,
while it decreases when the FWHM decreases.

\section{Conclusions}

In summary, we have computed the Raman
spectra of gated bilayer graphene, which is strongly influenced by the interaction with the environment.
We claim that by the analysis of the splitting of the $G$ mode in Raman measurements 
it is possible to estimate the amount of unintentional charges
coming from the atmosphere and from the substrate. 
Here we compare our calculations with the only experimental data available on the $G$ mode
splitting in bilayer graphene, and we give an estimate of the unintentional charges 
coming from the environment in this experiment.
In order to facilitate the comparison of new experimental results with our theoretical calculations, 
we provide as additional material a set of computed Raman shifts, FWHM, and
relative Raman intensities as a function of $n$,
for different values of $n_{top}$. \cite{files}


\begin{thebibliography}{35}
\bibitem[{fit()}]{fit}
\bibinfo{note}{In Ref.\cite{Pimenta-phonon} the authors fit the raw
  experimental data with Lorentzians of variable FWHM. Here we re-fit data from
  Ref.\cite{Pimenta-phonon} for all gate voltages with two Lorentzians of fixed
  FWHM equal to 7 ${\rm cm^{-1}}$, which is the value obtained in
  Ref.\cite{Pimenta-phonon} for large negative voltages, where the two peaks
  are well resolved. Our choice is justified by the fact that, according to our
  calculations, the EPC contribution to the FWHM is vanishing around $n^0$=
  -1.8 ${\rm 10^{13}cm^{-2}}$, for all considered $n_{top}$
  (Fig.\ref{shift-1}). Thus, the increase in broadening observed in
  Ref.\cite{Pimenta-phonon} for positive value of gate voltage is not due to an
  increase of the decay rate ($\gamma_i$) but to the presence of two
  contributions of fixed width in the Raman peak, close in frequency but not
  exactly superimposed. The root mean square error obtained with the two
  fitting procedures are comparable.}
\end{thebibliography}
\end{document}